\documentclass[%
 groupedaddress,
 aip,
 jmp,%
 amsmath,amssymb,
nofootinbib,
preprint,%
]{revtex4-1}

\usepackage{graphicx}
\usepackage{dcolumn}
\usepackage{bm}
\usepackage{color}

\usepackage{xcolor}

\begin{document}

\preprint{}

\title[Relativistic two-stream instability in low-density plasma] {
A relativistic two-stream instability in an extremely low-density plasma} 

\author{Shinji Koide}
\affiliation{Department of Physics, Kumamoto University, 
2-39-1, Kurokami, Kumamoto, 860-8555, JAPAN}
\email{koidesin@kumamoto-u.ac.jp}
\author{Masaaki Takahashi}
\affiliation{Department of Physics and Astronomy, Aichi University of Education,
Kariya, Aichi 448-8542, Japan}
\author{Rohta Takahashi}
\affiliation{National Institute of Technology (KOSEN), Tomakomai College,
443, Nishikioka, Tomakomai, Hokkaido, 059-1275, Japan}

\date{\today}

\begin{abstract}
A linear analysis based on two-fluid equations in the 
approximation of a cold plasma, wherein the plasma temperature is assumed to be zero, 
demonstrates that a two-stream instability occurs in all cases.
\textcolor{black}{However, if this were true, 
the drift motion of electrons in an
electric current over a wire would become unstable, inducing 
an oscillation 
in an electric circuit with ions bounded around specific positions.
To avoid this peculiar outcome, we must assume a warm plasma with
a finite temperature when discussing the criterion of instability.}
The two-stream instability \textcolor{black}{in warm plasmas} has typically been 
analyzed
using kinetic theory \textcolor{black}{to provide a general formula for the 
instability criterion from the distribution function of the plasma.}
\textcolor{black}{However, the criteria
based on kinetic theory do not have an easily applicable form.}
Here, we provide an easily applicable criterion for the instability
based on the two-fluid model at finite temperatures, 
\textcolor{black}{extensionally} 
in the framework of special relativity.
\textcolor{black}{This criterion is relevant for analyzing two-stream
instabilities in low-density plasmas in the universe and in Earth-based
experimental devices.}

\end{abstract}

\pacs{47.65.-d, 52.20.-j, 52.35.Qz, 97.60.Lf}
\maketitle

\maketitle

\section{Introduction} \label{sec2}

Investigations of the two-stream instability in plasmas have been ongoing for 
more than five decades.\cite{buneman59,krall86}
The research field was highly active before the 2000s. 
For instance,
the two-stream instability has been utilized as an elemental mechanism of
anomalous resistivity for magnetic reconnection,
which is crucial for energy release from magnetic
fields to plasma in solar flares and ionospheric phenomena,
like the aurora.\cite{papadopoulos77}
The two-stream instability was first demonstrated experimentally by
Pierce and Heibenstreit.\cite{pierce49}
Since then, numerous subsequent experimental investigations and
an exhaustive theoretical analysis of the two-stream instability have been conducted.\cite{krall86}
However, in the last two decades, publications on the two-stream instability
have been limited
because no new critical event related to the two-stream instability
has been found in astrophysical
or experimental plasmas.
Recently, we realized that the two-stream instability may solve
one issue in high-energy astrophysics
\textcolor{black}{in which relativistic effects are significant}.
Therefore, we require a \textcolor{black}{relativistic} criterion
that is readily applicable to evaluate the two-stream instability in this problem.
However, such an easily applicable criterion
has not yet to be demonstrated 
\textcolor{black}{within the relativistic framework}.

Various types of two-stream instabilities 
occur in plasmas with different particle species and conditions.
The two-stream instability of \textcolor{black}{ion-electron plasmas}
can be classified into two categories, the ``ion acoustic instability" and
the ``Buneman instability", depending on the relative velocity between the electron fluid 
and ion fluid (the electron fluid's drift velocity or drift velocity).\cite{lapuerta02}
\textcolor{black}{Hereafter, we will call an ion-electron plasma a ``normal plasma",
and an electron-positron plasma a ``pair plasma".}
When the drift velocity is large compared to the thermal velocity of the plasma,
the two-stream instability is known as the Buneman instability.\cite{buneman59}
When the drift velocity is modest, it is known as the ion acoustic instability.
\cite{jackson60,krall86}
A basic issue of great interest is which plasma conditions define the two-stream instability threshold. 
The ion acoustic instability has been explored \textcolor{black}{most often} 
using kinetic theory
with the Vlasov equation.\cite{fried61,kindel71,papadopoulos77,nakar11,hou15}
The ion acoustic instability is understood as the inverse process
of Landau damping\cite{landau46} between the ion fluid and electron fluid of the plasma.
The general form of the criterion for the ion acoustic instability 
is given by extending the Penrose criterion.\cite{penrose60}
The Penrose criterion is general but not transparent, and 
its application to astrophysical and experimental plasmas requires a numerical computation. \cite{lapuerta02}
For the Buneman instability, the two-fluid equations for a cold plasma
have been utilized, in which the plasma temperature is assumed to vanish.
\cite{buneman59,chen16,miyamoto87}
Correspondingly, the dispersion relation is obtained explicitly
using a linear analysis of two-fluid equations for the two-stream plasma. 
The dispersion relation
shows that the two-stream plasma is always unstable if the wave number is small enough.
This result is odd because if it were true, 
the drift motion of electrons in an
electric current via a wire would become unstable, causing an oscillation 
in an electric circuit, with ions bounded around specific positions.
When we consider a warm plasma whose temperature is not zero, the two-stream
instability should be suppressed at all wave numbers. 
\textcolor{black}{
In fact, Cordier, Grenier, and Guo succeeded in demonstrating a simple and general criterion
for the two-stream instability in the non-relativistic framework
\cite{cordier00} among a number of
investigations of instabilities in the nonrelativistic two-fluid model.
\cite{jao16,saleem05,blesson15,mohammadnejad19}
On the other hand, within the relativistic framework, a simple and general
criterion for the two-stream instability has not yet been found, despite having been investigated using relativistic kinetic theory
\cite{white85,hao09} and the relativistic two-fluid model.\cite{samuelsson11,haber16}}

\textcolor{black}{
In this paper, we present an easily applicable criterion for 
the two-stream instability
in a normal plasma within the framework of special relativity. 
The criterion is derived from an analysis of the linear
dispersion relation in terms of the special relativistic two-fluid equations 
with a finite temperature of the normal plasma.
The special relativistic effects are significant when the drift velocity
is close to the speed of light and/or the ion/electron temperature
is close to or greater than the electron rest mass energy.
This demonstrates that the two-stream instability occurs
if and only if the relativistic composition velocity of the sound velocities 
of the two fluids is less than the electron fluid's drift velocity.
This is a relativistic extension of the nonrelativistic criterion
given by Cordier, Grenier, and Guo. \cite{cordier00}}
Here, a noncollisional, unmagnetized, homogeneous,
normal plasma is considered. The criterion is so simple that it can be used
for any plasma process, whether astrophysical or on Earth.
It is worth noting
that when employing the two-fluid model, the inverse Landau damping
is disregarded,
and the instability treated by the criterion is the Buneman instability.
In a current sheet, the two-stream instability may play a key role,
particularly in regions of extremely low density. 
In the universe, such extremely low-density
regions are expected to exist near supermassive black holes in active regions,
from which relativistic astrophysical jets are ejected.

\section{The criterion of the two-stream instability
\label{deri2fluinst}}
\textcolor{black}{
\subsection{The dispersion relation with special-relativistic 
two-fluid equations}}

We derive the criterion of the two-stream instability of the normal plasma
\textcolor{black}{within the framework of special relativity}.
\textcolor{black}{For simplicity, we assume that the unperturbed plasma is uniform 
and charge-neutral,
with the ion fluid at rest and the electron fluid moving uniformly with velocity 
$v_0 > 0$
and Lorentz factor $\displaystyle \gamma_0 = (1 - v_0^2/c^2)^{-1/2}$ 
($c$ is the speed of light and the magnetic field is negligible)}.

The line element in Minkowski space-time $x^\mu = (ct, x^i)
=(ct, x, y, z)$ is given by
\begin{equation}
ds^2 = \eta_{\mu\nu} dx^\mu dx^\nu,
\end{equation}
where $\eta_{\mu\nu} = {\rm diag} (-1,1,1,1)$ 
is the Minkowski metric and $c$ is the speed of light.
The alphabetic index ($i, j, k$) runs from 1 to 3, while
the Greek index ($\mu, \nu$) runs from 0 to 3.
We utilize the special-relativistic two-fluid equations
with respect to the ion fluid and electron fluid and the inhomogeneous Maxwell equations.
The covariant form of the equation of continuity and the conservation of momentum and 
energy for the ion and electron fluids are given by
\begin{eqnarray}
\partial_\mu (n_\pm u_\pm^\mu ) &=& 0 ,
\label{eqcon_fluid} \\
\partial_\nu ( h_\pm u_\pm^\mu u_\pm^\nu  &+& p_\pm \eta^{\mu\nu} ) = f_\pm^\mu,
\label{covf_fluid}
\end{eqnarray}
where $n_\pm$, $p_\pm$, $h_\pm$, $u_\pm^\mu = (\gamma_\pm, u_\pm^i)$, and
$f_\pm^\mu$ are the proper number density,
pressure, proper enthalpy density, 4-velocity, and 4-Lorentz force density 
of the ion fluid (with subscript `$+$') and electron fluid (with subscript `$-$'), respectively.
We assume the ion and electron fluids are adiabatic with adiabatic indices
$\Gamma_+$ and $\Gamma_-$, respectively. Then 
$\displaystyle \frac{p_\pm}{n_\pm^{\Gamma_\pm}}$ is constant
and we have $h_\pm = n_\pm m_\pm c^2 + \frac{\Gamma_\pm}{\Gamma_\pm -1} p_\pm$,
where $m_\pm$ is the rest mass of the ion fluid
and the electron fluid, respectively.
The covariant form of Ampere's law and Gauss law with respect to the electric field is
\begin{equation}
\partial_\mu F^{\nu\mu} = \mu_0 J^\nu, 
\label{covf_field}
\end{equation}
where
$F_{\mu\nu}$, $J^\mu =(\rho_{\rm e} c, J^i) = e ( n_+ u_+^\mu -n_- u_-^\mu)
$, $\mu_0$, and $e$ are the field strength tensor, 
4-current density, magnetic permeability, and elementary charge, respectively.
The field strength tensor is related to the electric field $E_i$ and
the magnetic field $B^i$ as $F_{i0} = - F_{0i} = E_i$ 
and $F_{ij} = \epsilon_{ijk} B^k$, where $\epsilon_{ijk}$ is the Levi--Civita symbol.
The 4-Lorentz force densities on the ion and electron fluids are given by
\begin{equation}
f_\pm^\mu 
= \pm e n_\pm {F^\mu}_{\nu} u_\pm^\nu.
\end{equation}
In the case of zero magnetic field,
Eqs. (\ref{eqcon_fluid}), (\ref{covf_fluid}), and (\ref{covf_field}) yield
\begin{eqnarray}
&& \frac{\partial}{\partial t} (n_\pm \gamma_\pm) + 
\frac{\partial}{\partial x^k} (n_\pm \gamma_\pm v_\pm^k) = 0, \\
&& M_\pm \gamma_\pm^2  \left (\frac{\partial}{\partial t}
+ v_\pm^k  \frac{\partial}{\partial x^k} \right ) v_\pm^i 
= \pm e \gamma_\pm \left [ \textcolor{black}{E^i - \frac{1}{c^2} v^i v^j E_j } \right ]
- \frac{1}{n_\pm} \left (
\frac{\partial}{\partial x^i} + \frac{v^i}{c^2}  \frac{\partial}{\partial t}
\right ) p_\pm, \\
&& \frac{\partial}{\partial x^i} E^i = \rho_{\rm e} = e (n_+ \gamma_+ - n_- \gamma_-),
\end{eqnarray}
where $\displaystyle M_\pm = \frac{h_\pm}{n_\pm c^2}$ is the effective mass 
of the particle and
$v_\pm^k$ is the 3-velocity of the ion fluid and electron fluid.
For convenience, we employ a unit system where the electric permittivity 
is unity for a while ($\epsilon_0 = 1$).
We consider the perturbation of the velocities, Lorentz factors, electric field, densities,
pressures, enthalpy densities, effective masses, and charge density
on the equilibrium state of the ion and electron fluids and field:
$v_+^i = (\delta v_+, 0, 0)$,
$v_-^i = (v_0+ \delta v_-, 0, 0)$,
\textcolor{black}{$\gamma_+ = 1$, $\gamma_- = \gamma_0 + \delta \gamma_-$,}
$E^i = (\delta E, 0, 0)$,
$n_+ = \bar{n}_+ + \delta n_+$, 
$n_- =  \bar{n}_- + \delta n_-$, 
$p_\pm = \bar{p}_\pm + \delta p_\pm$, 
$h_\pm = \bar{h}_\pm + \delta h_\pm$,
$M_\pm = \bar{M}_\pm + \delta M_\pm$,
$\rho_{\rm e} = \bar{\rho}_{\rm e} + \delta \rho_{\rm e}$,
where the variables with the subscript bar express stationary and uniform states
and the variables with ``$\delta$" express the perturbations.
It is noteworthy that in the case with the large Lorentz factor $\gamma_0$, we have to
take $\delta v_- \ll v_0/\gamma_0^2$ \textcolor{black}{because $\delta \gamma_- =
\gamma_0^3 (v_0/c^2) \delta v_-$}.
When we assume the charge neutrality of the equilibrium state $\bar{\rho}_{\rm e} = 0$, 
we get $\bar{n}_+ = \bar{n}_- \gamma_0 \equiv n_0$.
We then have the following linearized equations:
\begin{eqnarray}
& \frac{\partial}{\partial t} \delta n_+
 + \frac{\partial}{\partial x} (n_0 \delta v_+ ) = 0,   \\
& \frac{\partial}{\partial t} \left (
\gamma_0 \delta n_- + n_0 \gamma_0^2 \frac{v_0}{c^2} \delta v_- \right )
+ \frac{\partial}{\partial x} (n_{0} \delta v_-) 
+ \frac{\partial}{\partial x} \left [ v_0 (\gamma_0 \delta n_-
+ n_0 \gamma_0^2 \frac{v_0}{c^2} \delta v_- ) \right ] = 0, 
\label{lineqconele} \\
& \bar{M}_+ \frac{\partial}{\partial t} \delta v_+ 
= e \delta E - \frac{1}{n_0} \frac{\partial}{\partial x} \delta p_+ , \\
& \bar{M}_- \gamma_0^3 \left (\frac{\partial}{\partial t} 
+ v_0 \frac{\partial}{\partial x} \right ) \delta v_-
= - e \delta E - \frac{1}{n_0} \frac{\partial}{\partial x} \delta p_- , 
\label{lineqmotele} \\
& \frac{\partial}{\partial x} \delta E = e \left ( \delta n_+ - \gamma_0 \delta n_-
- n_0 \gamma_0^2 \frac{v_0}{c^2} \delta v_- \right), 
\label{lineqgaus} \\
& \frac{\delta p_+}{P_+}  = \Gamma_+ \frac{\delta n_+}{n_0} , 
 \frac{\delta p_-}{P_-}  = \Gamma_- \frac{\delta n_-}{\bar{n}_-},
\end{eqnarray}
where we use $P_\pm \equiv \bar{p}_\pm$ and we assume 
$\displaystyle \partial_i =\left ( \frac{\partial}{\partial x}, 0, 0\right )$.
When we introduce $\bar{M}^\ast_- \equiv \gamma_0^3 \bar{M}_-$, $\delta n^\backprime_- \equiv
\gamma_0 \delta n_- + n_0 \gamma_0^2 v_0 \delta v_-/c^2$, and
$\delta p_-^\backprime \equiv \Gamma_- \frac{P_-}{n_-} \gamma_0^3 \delta n^\backprime_-$,
Eqs. (\ref{lineqconele}), (\ref{lineqmotele}), (\ref{lineqgaus}) yield
\begin{eqnarray}
& \displaystyle
\frac{\partial}{\partial t} \delta n^\backprime_-
+ \frac{\partial}{\partial x} (n_{0} \delta v_-) 
+ \frac{\partial}{\partial x} (v_0 \delta n^\backprime_-) = 0, \\
& \displaystyle
\bar{M}^\ast_- \left (\frac{\partial}{\partial t} 
+ v_0 \frac{\partial}{\partial x} \right ) \delta v_-
= - e \delta E - \frac{\gamma_0^2}{n_0} \left (
 \frac{\partial}{\partial x} + \frac{v_0}{c^2} \frac{\partial}{\partial t} \right ) 
\delta p_-^\backprime , \\
& \displaystyle 
\frac{\partial}{\partial x} \delta E = e ( \delta n_+ - \delta n^\backprime_-).
\end{eqnarray}
Assuming that the perturbations vary as $\exp (i k x - i \omega t)$, we obtain
\begin{eqnarray}
- i \omega \delta n_+ + i k (n_0 \delta v_+) = 0,
&\quad& - i \omega \delta n^\backprime_- + i k (n_{0} \delta v_-) 
+ i k v_0 \delta n_-^\backprime = 0, 
\label{expnexpn1} \\
- i \omega \bar{M}_+ \delta v_+ = e \delta E - i \frac{1}{n_0} k \delta p_+ , 
&\quad& (- i \omega + i k v_0) \bar{M}^\ast_- \delta v_- 
= - e \delta E - i \frac{1}{n_{0}} k
\left ( 1 - \frac{v_0 \omega}{c^2 k} \right )^2 \delta p_-^\backprime , \\
i k \cdot \delta E = e ( \delta n_+ - \delta n^\backprime_-), \\
 \delta p_+ = \Gamma_+ \frac{P_+}{n_0} \delta n_+ = \bar{M}_+ c_{\rm i}^2 \delta n_+, 
&& 
\delta p_-^\backprime 
= \bar{M}_-^\ast c_{\rm e}^2 \delta n^\backprime_-, 
\label{expnexpn4}
\end{eqnarray}
where
$\displaystyle c_{\rm i} = \sqrt{\frac{\Gamma_+ P_+}{n_0 \bar{M}_+}}
= c \sqrt{\frac{\Gamma_+ P_+}{\bar{h}_+}}$ and 
$\displaystyle c_{\rm e} = \sqrt{\frac{\Gamma_- P_- \gamma_0^3}{n_- \bar{M}_-^\ast}}
= c \sqrt{\frac{\Gamma_- P_-}{\bar{h}_-}}$ 
are the sound speed in the ion and electron fluids, respectively.
\textcolor{black}{
Note that the sound velocity of the plasma is given by $\displaystyle 
c_{\rm s} = \sqrt{\frac{c_{\rm i}^2 + \epsilon c_{\rm e}^2}{1 + \epsilon}}$,
which is called the ion acoustic velocity.
In the case of a normal plasma, the ion acoustic velocity $c_{\rm s}$
is approximately equal to $c_{\rm i}$ 
as long as
$c_{\rm e}$ is not much larger than $c_{\rm i}$
because $\epsilon \ll 1$.}
Eventually, we obtain the dispersion relation equation with Eqs.
(\ref{expnexpn1})--(\ref{expnexpn4}),
\begin{equation}
\frac{\epsilon}{\omega^2 - c_{\rm i}^2 k^2}
+ \frac{1}{(\omega - k v_0)^2 - c_{\rm e}^2 (k - v_0 \omega/c^2)^2}
= \frac{1}{\omega_{\rm pe}^2}
\label{dispersiontwofluid}
\end{equation}
where $\epsilon = \bar{M}^\ast_-/\bar{M}_+ = \gamma_0^3 \bar{M}_-/\bar{M}_+$ 
is the modified electron-ion
mass ratio 
and $\displaystyle \omega_{\rm pe} = \sqrt{\frac{n_0 e^2}{\bar{M}^\ast_-}}
= c \sqrt{\frac{n_0^2 e^2}{\gamma_0^4 \bar{h}_-}}$
is the relativistic electron plasma frequency.

\textcolor{black}{
\subsection{Derivation of the criterion of the two-steam
instability and further analysis}
We analyze the dispersion relation (\ref{dispersiontwofluid}) to derive the criterion
of the two-stream instability.}
For convenience, we use the inverse of the modified electron plasma frequency 
$\displaystyle \omega_{\rm pe}^{-1}$ as the unit of time.
Equation (\ref{dispersiontwofluid}) yields 
\begin{equation}
F(\omega) \equiv  
\frac{\epsilon}{\omega^2 - c_{\rm i}^2 k^2}
+ \frac{1}{(\omega - v_0 k)^2 - c_{\rm e}^2 (k - v_0 \omega/c^2)^2}  = 1,
\label{dispersiontwofluidnormalizd}
\end{equation}
which is a quartic algebraic equation with respect to $\omega$.
If the number of real solutions of Eq. (\ref{dispersiontwofluidnormalizd}) 
is less than three, Eq. (\ref{dispersiontwofluidnormalizd}) 
has at least one complex number solution, which represents the instability.
As shown below, if and only if $\displaystyle
\frac{c_{\rm i} + c_{\rm e}}{1+c_{\rm i} c_{\rm e}/c^2} < v_0$, 
$F(\omega) = 1$ has only
two real number solutions. We determine the criterion of the two-stream instability
as
\begin{equation}
c_{\rm i} \oplus c_{\rm e} 
= \frac{c_{\rm i} + c_{\rm e}}{1+c_{\rm i} c_{\rm e}/c^2} < v_0.
\label{unstabcond}
\end{equation}
Hereafter, we use the sign ``$\oplus$" to describe the relativistic velocity composite 
law 
\textcolor{black}{so that $\displaystyle a \oplus b \equiv \frac{a + b}{1 + ab/c^2}$, 
where $a$ and $b$ are arbitrary velocities.
Furthermore, we introduce the sign ``$\ominus$" 
so that $\displaystyle a \ominus b \equiv \frac{a - b}{1 - ab/c^2}$.}
On the other hand, when $c_{\rm i} \oplus c_{\rm e} \ge v_0$, $F(\omega) = 1$ has 
four real number solutions, which represent a stable oscillation or wave. 
We have a stable condition
for the two-stream instability, $c_{\rm i} \oplus c_{\rm e} \ge v_0$.

\begin{figure}
\begin{minipage}{1.0 \hsize}
\begin{center}
\includegraphics[scale=0.7]{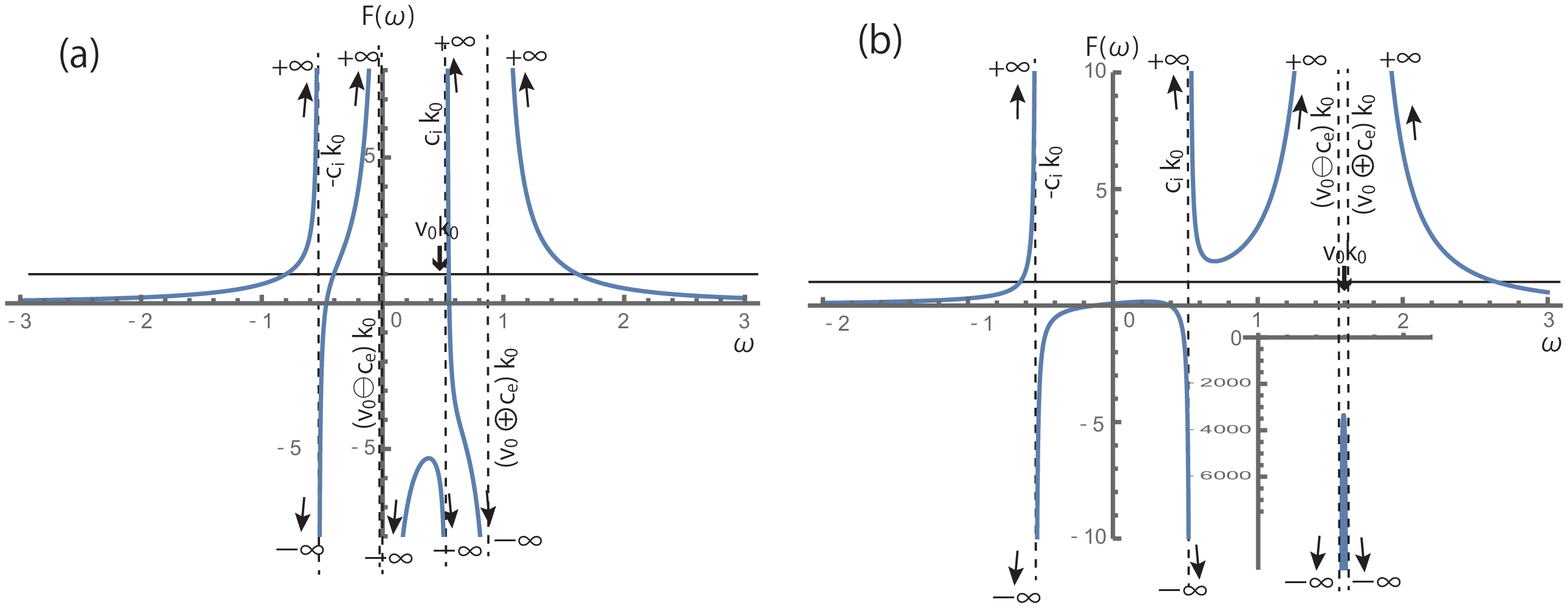}
\end{center}
\caption{A plot of $F(\omega)$ in the relativistic cases of 
$c_{\rm i} \oplus c_{\rm e} \ge v_0$ with
$v_0 \textcolor{black}{ =0.984c} = 1$, $c_{\rm i} = c_{\rm e} \textcolor{black}{ = 0.984 c} = 1$ [panel (a), case (ii)]
and of $c_{\rm i} \oplus c_{\rm e} < v_0$ with
$v_0 \textcolor{black}{ = 0.984c} = 3$, $c_{\rm i} = c_{\rm e} \textcolor{black}{ = 0.328 c} = 1$ [panel (b), case (i)], where 
$\gamma_0^3 =183.6$, that is, $\epsilon = 1/10$ and $k = 0.5313$.
The cross point on the curve of $F(\omega)$ and the line $F=1$ 
give the dispersion relation. In the case of (a) $c_{\rm i} \oplus c_{\rm e} \ge v_0$
[case (ii)], 
all four solutions of the dispersion relation are real numbers. 
In the case of (b) $c_{\rm i} \oplus c_{\rm e} < v_0$ [case (i)], 
only two solutions of the dispersion
relation are real numbers and two complex solutions exist. 
Not that in panel (b), $F$ between 
$\displaystyle \omega = 
(v_0 \ominus c_{\rm e}) k_0$ and $\omega = (v_0 \oplus c_{\rm e}) k_0$
is plotted in a separate box because its scale is quite different from the 
other range.
\label{fig1}}
\end{minipage}
\end{figure}

\begin{figure}
\begin{minipage}{1.0 \hsize}
\begin{center}
\includegraphics[scale=0.7]{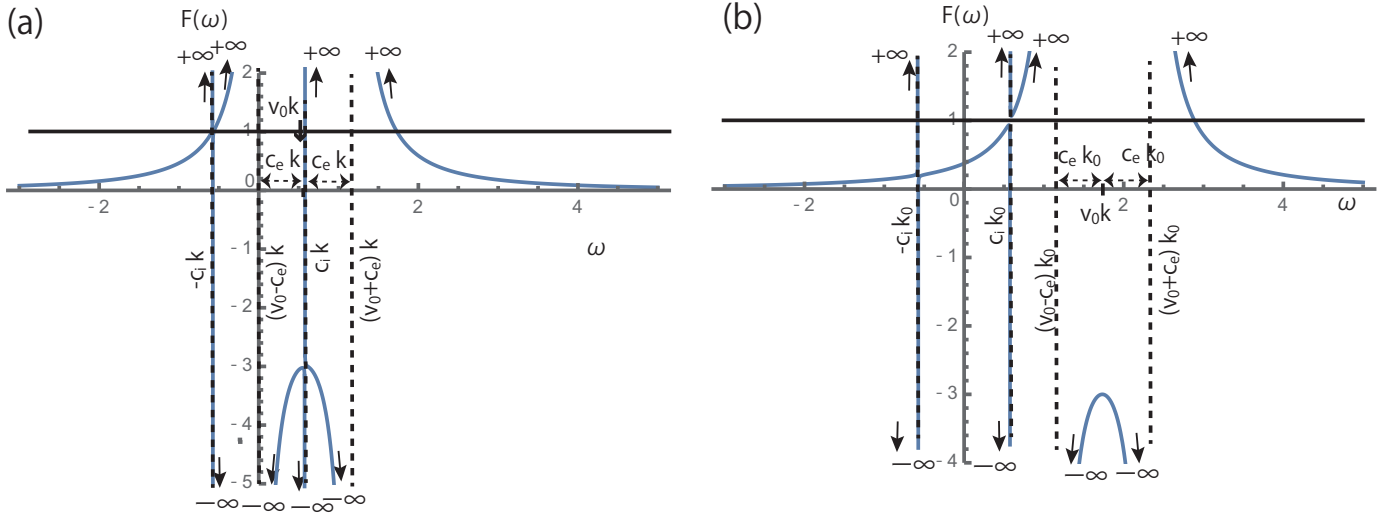}
\end{center}
\caption{As Fig. \ref{fig1}, but for 
$F(\omega)$ in the non-relativistic cases of
$c_{\rm i} + c_{\rm e} \ge v_0$ with 
$v_0 =  1$, $c_{\rm i} = c_{\rm e} =  1$ [panel (a), case (ii)]
and of $c_{\rm i} + c_{\rm e} < v_0$ with
$v_0 =  3$, $c_{\rm i} = c_{\rm e} =  1$ [panel (b), case (i)], where
$\epsilon = 1/1836$ ($\gamma_0 = 1$) and $k=0.5313$. 
In the non-relativistic case, we can simply write 
$c_{\rm i} \oplus c_{\rm e} =c_{\rm i} + c_{\rm e} $
and $c_{\rm i} \ominus c_{\rm e} =c_{\rm i} - c_{\rm e}$.
\label{fig2}}
\end{minipage}
\end{figure}

\begin{figure}
\begin{minipage}{1.0 \hsize}
\begin{center}
\includegraphics[scale=0.5]{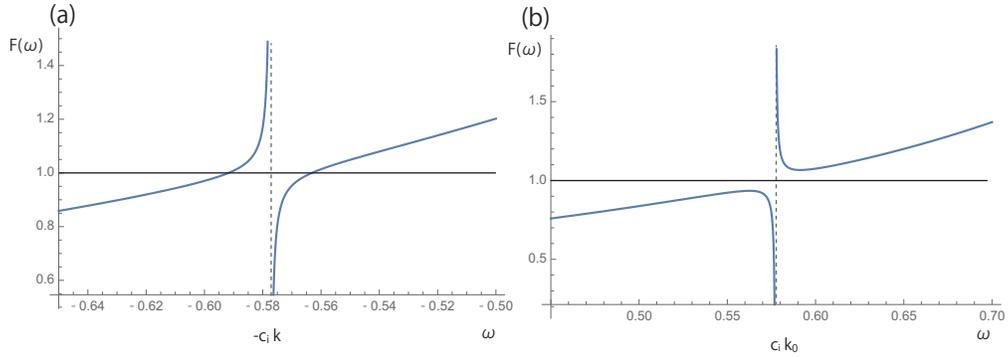}
\end{center}
\caption{A zoomed-in version of the plot of $F(\omega)$ in the non-relativistic cases of
$c_{\rm i} + c_{\rm e} \ge v_0$ with $v_0 =1$, $c_{\rm i} = c_{\rm e} = 1$ 
[case (ii)] around $\omega = - c_{\rm i} k = - 0.577$ [panel (a)] 
and of $c_{\rm i} + c_{\rm e} < v_0$ with $v_0 =3$, $c_{\rm i} = c_{\rm e} = 1$ 
 around $\omega = c_{\rm i} k = 0.577$ 
[panel (b)], where $\epsilon = 1/1836$ and $k=1/\sqrt{3}=0.577$. 
\label{fig3}}
\end{minipage}
\end{figure}

\textcolor{black}{$F(\omega)$ has four singular points, 
$\omega = \pm c_{\rm i} k$, $(v_0 \ominus c_{\rm e})k$, $(v_0 \oplus c_{\rm e})k$, 
where 
$\displaystyle v_0 \ominus c_{\rm e} \equiv 
\frac{v_0 - c_{\rm e}}{1- v_0 v_{\rm e}/c^2}$.}
The behaviors of $F(\omega)$ at the singular points are as follows,
\begin{eqnarray}
\lim_{\omega \rightarrow - c_{\rm i} k -0} F(\omega) = \infty ,& &
\lim_{\omega \rightarrow - c_{\rm i} k +0} F(\omega) = - \infty , \\
\lim_{\omega \rightarrow c_{\rm i} k -0} F(\omega) = - \infty ,& &
\lim_{\omega \rightarrow c_{\rm i} k +0} F(\omega) =  \infty , \\
\lim_{\omega \rightarrow (v_0 \ominus c_{\rm e})k-0} F(\omega) =  \infty ,& &
\lim_{\omega \rightarrow (v_0 \ominus c_{\rm e})k+0} F(\omega) = - \infty , \\
\lim_{\omega \rightarrow (v_0 \oplus c_{\rm e})k-0} F(\omega) = - \infty ,& &
\lim_{\omega \rightarrow (v_0 \oplus c_{\rm e})k+0} F(\omega) = \infty.
\end{eqnarray}
There are \textcolor{black}{four} kinds of patterns in the relationship of the 
four singular points:
\textcolor{black}{
(i) $- c_{\rm i} k < c_{\rm i} k < (v_0 \ominus c_{\rm e}) k < (v_0 \oplus c_{\rm e}) k$,
(ii) $- c_{\rm i} k <  (v_0 \ominus c_{\rm e}) k \le c_{\rm i} k <(v_0 \oplus c_{\rm e})k$,
(iii) $- c_{\rm i}k <  (v_0 \ominus c_{\rm e}) k <(v_0 \oplus c_{\rm e})k \le c_{\rm i} k$,
(iv) $(v_0 \ominus c_{\rm e})k \le - c_{\rm i}k < c_{\rm i}k < (v_0 \oplus c_{\rm e})k$
(see Table \ref{table2mt}).
The topologies of the profiles of $F(\omega)$ in case (i) and case (ii)
are drastically different and the numbers of real solutions for $F(\omega) = 1$
are different. The profiles of $F(\omega)$ in the cases (ii), (iii), and (iv)
are topologically identical and the numbers of the real solutions of
$F(\omega) = 1$ are the same.
The condition for case (i) is given by
(i) $c_{\rm i} < v_0 \ominus c_{\rm e}$ and
the condition of cases (ii), (iii), (iv) can be summarized as
(ii--iv) $v_0 \ominus c_{\rm e} \le c_{\rm i}$ because $v_0$ is positive.
The conditions for cases (i) and (ii--iv) are written as
(i) $c_{\rm i} \oplus c_{\rm e} < v_0$ and
(ii--iv) $v_0 \le c_{\rm i} \oplus c_{\rm e}$, respectively.}

\begin{table}
\begin{tabular}[b]{|c|c|c|c|c|c|c|c|c|c|c|c|} \hline
(i) & $-\infty$ & $\cdots$& $-c_{\rm i} k$ & $\cdots$ & $c_{\rm i} k$ & $\cdots$ & $(v_0 \ominus c_{\rm e}) k$ & $\cdots$ & $(v_0 \oplus c_{\rm e}) k$  & $\cdots$ &  $+\infty$ \\
 & & & -- \hspace{1em} + & & -- \hspace{1em} + & & -- \hspace{1em} + & & -- \hspace{1em} + & & \\ \hline
$F(\omega)$ & 0 &$\nearrow$ & $+\infty$ $-\infty$ &  $\nearrow$ $\searrow$ & $-\infty$ $+\infty$ & $\searrow$ $\nearrow$ & $+\infty$ $-\infty$ & $\nearrow$ $\searrow$ & $-\infty$ $+\infty$ & $\searrow$ & 0 \\ \hline \hline
(ii) & $-\infty$ & $\cdots$& $-c_{\rm i} k$ & $\cdots$  & $(v_0 \ominus c_{\rm e}) k$ & $\cdots$ & $c_{\rm i} k$ & $\cdots$ & $(v_0 \oplus c_{\rm e}) k$  & $\cdots$ &  $+\infty$  \\ \hline
(iii) & $-\infty$ & $\cdots$& $-c_{\rm i} k$ & $\cdots$  & $(v_0 \ominus c_{\rm e}) k$ & $\cdots$ & $(v_0 \oplus c_{\rm e}) k$  & $\cdots$ & $c_{\rm i} k$ & $\cdots$ &  $+\infty$  \\ \hline
(iv) & $-\infty$ & $\cdots$& $(v_0 \ominus c_{\rm e}) k$ & $\cdots$  & $-c_{\rm i} k$ & $\cdots$ & $c_{\rm i} k$ & $\cdots$ & $(v_0 \oplus c_{\rm e}) k$  & $\cdots$ &  $+\infty$  \\ \hline
$F(\omega)$ & 0 &$\nearrow$ & $+\infty$ $-\infty$ &  $\nearrow$ & $+\infty$ $-\infty$ & $\nearrow$ $\searrow$ & $-\infty$ $+\infty$ & $\searrow$ & $-\infty$ $+\infty$ & $\searrow$ & 0  \\ \hline
\end{tabular}
\caption{A plot of the singular points of $F(\omega)$.
\textcolor{black}{In the region of ``$\nearrow$ $\searrow$", $F(\omega)$
has its maximum value, and $F(\omega)$ in the region ``$\searrow$ $\nearrow$"
has its minimum value. ``$\nearrow$" represents a monotonic increase and 
``$\searrow$" represents a monotonic decrease.
In case (i), the number of the real number solutions
of $F=1$ is two for a small enough wave number.}
\label{table2mt}}
\end{table}

First, \textcolor{black}{heuristically,} we consider the relativistic case of 
$\gamma_0^3 = 183.6$, that is, $\epsilon = 1/10$, because 
in the non-relativistic case of $\epsilon = 1/1836$,
the function $F(w)$ has a very fine structure that is difficult to understand.
\textcolor{black}{
Hereafter, we \textcolor{black}{treat} $v_0$ and $\gamma_0$ as model-free parameters so that the light speed
\textcolor{black}{is} $\displaystyle c = \frac{\gamma_0}{\sqrt{\gamma_0^2 -1}} v_0 \ge v_0$.}

When $c_{\rm i} \oplus c_{\rm e} \ge v_0$ [case (ii)], the profile of $F(\omega)$
becomes as shown in Fig. \ref{fig1} (a) for any $k$, where we set
$\epsilon = 1/10$, $c_{\rm i}=c_{\rm e} \textcolor{black}{ =0.984c}=1$, $v_0 \textcolor{black}{ =0.984c} = 1$ 
\textcolor{black}{($c=1.016$)}, and $k=0.5313$.
\textcolor{black}{Throughout this paper, for the numerical calculation, we normalize the velocity
by $c_{\rm i}$. In the case of a normal plasma, the normalized value of the 
velocity is approximately the ion Mach number 
as long as $c_{\rm e}$ is not much larger than $c_{\rm i}$,
because $c_{\rm i}$ is approximately the ion acoustic velocity $c_{\rm s}$.}
\textcolor{black}{
It is found that we always have four real solutions of $F(\omega)=1$,
because there are four separate lines connecting the line 
$F=0$ ($F \longrightarrow 0$) and the region $F \longrightarrow \infty$.
In the case of $c_{\rm i} \oplus c_{\rm e} \ge v_0$,
the system is stable.} This holds for any $\epsilon$ as shown in
Fig. \ref{fig2} (a) for the case of $\epsilon = 1/1836$ 
\textcolor{black}{($\gamma_0=1, c \longrightarrow \infty$)}.
We show the detail of the line of $F(\omega)$ around $\omega = - c_{\rm i} k$
in Fig. \ref{fig3} (a), which clearly indicates two real solutions exist 
around $\omega = - c_{\rm i} k$.
\textcolor{black}{It is noted that the profiles of $F(\omega)$ in cases
(ii), (iii), and (iv) are similar, and then the equation $F(\omega) = 1$ has
four real solutions, not only in case (ii) but also in cases (iii) and (iv).}

On the other hand, in the case of $c_{\rm i} \oplus c_{\rm e} < v_0$ [case (i)],
the topological profile of $F(\omega)$ changes drastically.
We plot $F(\omega)$ in the case of $\epsilon = 1/10$ ($\gamma_0^3 = 183.6$), 
$c_{\rm i}=c_{\rm e} \textcolor{black}{ =0.328c}=1$, and $v_0 \textcolor{black}{ = 0.984c}= 3$ \textcolor{black}{($c=3.048$)} with
\begin{equation}
k = k_0(v_0, c_{\rm i}, c_{\rm e}) \equiv \frac{1}{\sqrt{(c_{\rm i} -v_0)^2 -c_{\rm e}^2
(1 - v_0 c_{\rm i}/c^2)^2}}
= \frac{\omega_{\rm pe}}{\sqrt{(c_{\rm i} -v_0)^2 -c_{\rm e}^2(1 - v_0 c_{\rm i}/c^2)^2}},
\label{eq_k0}
\end{equation}
in Fig. \ref{fig1} (b). Here, on the right-hand side of Eq. (\ref{eq_k0}), 
the unit of time is recovered from the specified unit of time $\omega_{\rm pe}^{-1}$.
\textcolor{black}{
It is found that $F(\omega)=1$ has only two real solutions because there are only two separate lines connecting the line 
$F=0$ ($F \longrightarrow 0$) and the region $F \longrightarrow \infty$, and the
other lines never touch or cross the line $F=1$.
The other two solutions are complex, which indicates that the two fluids are unstable.}
This instability is recognized as the ``two-stream instability".\cite{chen16}

In the non-relativistic case of $\epsilon = 1/1836$
\textcolor{black}{($\gamma_0 =1$, $c \longrightarrow \infty$)},
$c_{\rm i}=1$, $c_{\rm e}=1$, $v_0 = 3$ with $k=k_0(v_0, c_{\rm i}, c_{\rm e})$,
we plot the curve of $F(\omega)$ in Fig. \ref{fig2} (b). 
$F(\omega)=1$ has only two real solutions.
This is because the curve $F(\omega)$ never passes $\omega = c_{\rm i} k$,
as shown in Fig. \ref{fig3} (b).
The other two solutions are complex, which indicates that the two-fluid system is unstable. 
We then obtain the criterion
of the two-stream instability, $c_{\rm i} + c_{\rm e} 
= c_{\rm i} \oplus c_{\rm e} < v_0$
for any positive $\epsilon$, in the non-relativistic case.
\textcolor{black}{In summary, the criterion of the two-stream instability is given 
by Eq. (\ref{unstabcond}) in the special relativistic framework.}

\textcolor{black}{To investigate the range of the wave number of the unstable mode 
in case (i)},
we write $F(\omega)$ with $k=k_0$ as $F_0(\omega)$.
We call the maximum value of
$F_0(\omega)$ in the region $- c_{\rm i} k < \omega < c_{\rm i} k$ 
$F_{\rm 0,i}^{\rm max}$
and the minimum value of $F_0(\omega)$ in the region $c_{\rm i} k < \omega < (v_0
\oplus c_{\rm e}) k$ $F_{\rm 0,b}^{\rm min}$. The wave number of the unstable
mode is given by $k_0 \sqrt{\max(0, F_{\rm 0,i}^{\rm max})} < k 
< k_0 \sqrt{F_{\rm 0,b}^{\rm min}}$.
When we consider the imaginary case with $\epsilon = 1$ corresponding 
a non-relativistic pair plasma (an electron-positron plasma), 
the mode of $0 < k < k_0$ is unstable because $F_{\rm 0,i}^{\rm max} < 0$ 
and $F_{\rm 0,b}^{\rm min} > 1$.
On the other hand, in the non-relativistic realistic case of a normal plasma ($\epsilon 
= 1/1836 \ll 1$), the unstable mode is restricted to $k \approx k_0$
because $F_{\rm 0,i}^{\rm max} \approx F_{\rm 0, b}^{\rm min}$.

It is noteworthy that in the non-relativistic case of $\epsilon = 1/1836$,
$c_{\rm i}=1$, $c_{\rm e}=1$, $v_0 = 3$ 
\textcolor{black}{($\gamma_0 = 1$, $c \longrightarrow \infty$)}, we find the
curve of $F(\omega)$ with $k \ge 1.04 k_0$ or $k \le 0.96 k_0$
crosses the line $F = 1$ around $\omega \sim c_{\rm i} k$,
while the curve of $F(\omega)$ with $k \le 1.03 k_0$ and $k \ge 0.97 k_0$
never crosses the line $F = 1$ around $\omega \sim c_{\rm i} k$.
Then, the equation $F(\omega) = 1$ has four real solutions
and we find that the mode with with $k \ge 1.04 k_0$ or $k \le 0.96 k_0$
is stable, while the mode with with $k \le 1.03 k_0$ and $k \ge 0.97 k_0$
is unstable. 
Eventually, we conclude that, in the case of a normal plasma ($\epsilon = 1/1836 \ll 1$), 
the wave number of the unstable mode can be fixed to approximately
$k \cong k_0(v_0, c_{\rm i}, c_{\rm e})$.
In such a case, the difference between the complex solution of $F(\omega)=1$
and $\omega = b$ is infinitesimally small and we obtain the approximate solution
of $\omega = b + ip$, where $\displaystyle p = \pm \frac{1}{2 k_0}
\sqrt{\frac{\epsilon}{c_{\rm i} (v_0 - c_{\rm e})}}$.
We then find the growth rate of the two-stream instability 
\begin{equation}
p_{\rm gr} = \frac{1}{2} 
\sqrt{\epsilon \frac{(v_0-c_{\rm i})^2 - c_{\rm e}^2}{c_{\rm i} (v_0 - c_{\rm e})}} \omega_{\rm pe}
\end{equation}
in the non-relativistic case,
where we recover the dimension of time.




This linear analysis yields the criterion of the two-stream instability
in Eq. (\ref{unstabcond}). 
In an extremely low-density plasma, the current may become unstable due to 
the two-stream instability. This instability may be important for the plasma 
supply mechanism of relativistic jets from AGNs, as shown in the next section.


\section{Discussion \label{sec:discussion}}


In this study, we have presented a simple criterion for the two-stream instability
within the relativistic framework: the instability arises
if and only if $c_{\rm i} \oplus c_{\rm e} < v_0$.
This criterion is derived from the two-fluid equations for a plasma with a
finite temperature.
\textcolor{black}{
The two-stream instability with this criterion is identical to the
Buneman instability because the electron drift velocity is large compared
to the plasma thermal velocity.}

\textcolor{black}{
It is worth noting that Eq. (\ref{unstabcond}) can be written as follows
in order to determine
the relativistic effect on the two-stream instability criterion:
\begin{equation}
\frac{\gamma_{\rm i} \gamma_{\rm e}}{\gamma_{\rm 0}} (c_{\rm i} + c_{\rm e}) < v_0,
\label{mainformulae}
\end{equation}
where $\gamma_{\rm i} = (1-c_{\rm i}^2/c^2)^{-1/2}$ 
and $\gamma_{\rm e} = (1-c_{\rm e}^2/c^2)^{-1/2}$ 
are the Lorentz factors of the sound velocities in the
ion and electron fluids. Equation (\ref{mainformulae}) 
clearly indicates that
the thermal relativistic effect suppresses the instability, while
the relativistic effect of the electron drift motion enhances the instability.
}

\textcolor{black}{
This simple criterion for the instability described by Eq. 
(\ref{mainformulae}) applies to the plasma source
region of a relativistic jet close to a spinning supermassive black hole, such as
the active galactic nucleus (AGN) of the elliptical giant galaxy M87.
%
In some regions of the universe, such as at the pole of a black hole 
of an AGN like M87, \cite{mckinney09,eht19,porth19}
it is anticipated that plasmas with very low densities and strong magnetic fields exist. 
\textcolor{black}{Here, a strong magnetic field indicates that
the magnetic field energy is greater than the rest mass energy density
of the plasma.}
A number of numerical simulations of general relativistic magnetohydrodynamics
with zero electric resistivity (ideal GRMHD)
imply the presence of a 
very strong magnetic field in a foot-point region of the relativistic jet,
and the magnetic field lines along the relativistic jet are anchored
to the horizon of the black hole. Near the horizon,
the plasma in the magnetic flux
tube anchored to the horizon falls into the black hole.}
Because the magnetic field flux is supplied by an accretion disk
and the magnetic field in an accretion disk is turbulent, a large
antiparallel magnetic field is predicted to exist in the foot-point region of the jet.
At the boundary of the jet and the inflow toward the black hole,
the plasma density decreases infinitesimally because magnetic surfaces
disturb the plasma supply
from outside the region to the jet as a result of its frozen-in state.
If the plasma density decreases sufficiently, the drift velocity of the
electron fluid increases in the current sheet,
sustaining the antiparallel magnetic field. In such a scenario, 
the drift velocity of the electron fluid exceeds the threshold of
the two-stream instability, causing the instability to disrupt
the current sheet and annihilate
the magnetic field in the jet foot-point region. The annihilation of the
magnetic field in this region causes the plasma from outside the region to
rush in and \textcolor{black}{supply the plasma to the jet}.
\textcolor{black}{This is the first mechanism proposed to explain a normal plasma
supply to the relativistic jet ejected from an AGN, whereas a variety of}
pair plasma supply mechanisms have previously been described 
in the strong magnetic field region close to spinning supermassive black holes.
\cite{blandford77,levinson18,toma20}
Here, we have demonstrated that the two-stream instability in a plasma of 
extremely low density could transport
the normal plasma and magnetic field of the disk to the plasma 
source region of the jet. 
Once the plasma and magnetic field are supplied to the jet-forming region, 
the mechanism seen in ideal GRMHD simulations 
\cite{gammie03,mizuno04,koide06,nagataki09,mckinney06,mckinney09,porth19,eht19,eht22} 
and the general relativistic analytical theories of 
steady outflow \cite{takahashi90,hirotani92}
can work to produce a relativistic jet composed of normal plasma.
On the contrary, for the elemental process
due to the two-stream instability, 
we can use the special relativistic criterion even for plasma near the supermassive 
black hole,
\textcolor{black}{because the two-stream instability is a local phenomenon and we can neglect
the tidal force since the characteristic length of the instability $l_{\rm c}$ is
less than $2 \pi f_{\rm c} \lambda_{\rm D}$ 
from Eq. (\ref{eq_k0}) 
and $\lambda_{\rm D}$ is much smaller than
the gravitational radius of the supermassive black hole $r_{\rm H}$. 
Here, $\lambda_{\rm D}$ is the Debye length 
$\displaystyle \lambda_{\rm D} \equiv \frac{c_{\rm e}}{\omega_{pe}}$
and $f_{\rm c}$ is defined by $v_0 \ominus c_{\rm i} = f_{\rm c} c_{\rm e}$, where $f_{\rm c}$ is greater than
unity but not much greater.
We can then choose
local inertial frame coordinates such that no gravitational field appears
according to the equivalence principle.}

\begin{figure}
\begin{center}
\includegraphics[scale=0.8]{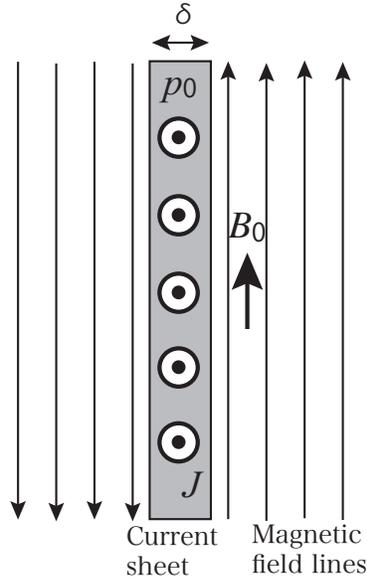}
\end{center}
\caption{ \textcolor{black}{
A schematic picture of the current sheet in the antiparallel magnetic field.}
\label{currentsheet}}
\end{figure}

Let us now consider the criterion of 
\textcolor{black}{the current disruption
in the low-density region at the pole of the black hole in an AGN,
for example, at the center of M87.
}
We assume an equilibrium condition with respect to the current sheet between the
antiparallel magnetic fields and the pressure of the normal plasma (Fig. \ref{currentsheet}),
\begin{align}
p_0 &= \frac{B_0^2}{2 \mu_0} , &
\mu_0 J \delta &= 2 B_0, &
J & = n e \gamma_0 v_0 = e n_0 v_0 , \label{cricurdiseqi}\\
p_0 &= P_+ + P_-,  &
P_+ & = n_+ T_+ = n_0 T , &
P_- & = n_- T_- = \frac{\xi}{\gamma_0} n_0 T 
\label{cricurdispe},
\end{align}
where $\delta$ is the width of the current sheet.
\textcolor{black}{
Here, we assume the pressure outside of the current sheet is negligibly small.
The magnetic field outside of the current sheet is denoted
$B$, and the particle number density, total pressure, ion pressure, electron pressure, 
current density, ion temperature, and electron temperature in the current sheet are denoted
$n_0$, $p_0$, $P_+$, $P_-$, $J$, $T_+ = T$, $T_- = \xi T$
($\xi$ is a constant parameter)}.
\textcolor{black}{Using Eqs. (\ref{cricurdiseqi})--(\ref{cricurdispe}),} we have
\textcolor{black}{
\begin{align}
P_+ & = \frac{\gamma_0}{\xi + \gamma_0} \frac{B_0^2}{2 \mu_0},
\label{cricurdisp0} 
P_-  = \frac{\xi}{\xi + \gamma_0} \frac{B_0^2}{2 \mu_0}, \\
v_0 & = \frac{2 B_0}{\mu_0 e n_0 \delta}
= 2 \sqrt{\frac{m \sigma c^2}{\mu_0e^2 n_0 \delta^2}},
\gamma_0 = \left ( 1 - \frac{4 m \sigma}{\mu_0 e^2 n_0 \delta^2} \right )^{-1/2},
\label{cricurdisce}
\end{align}
where $\sigma$ is the magnetization parameter defined by
$\displaystyle \sigma = \frac{B_0^2}{m n_0 c^2}$ with $m = m_+ + m_-$.
}
Equations (\ref{mainformulae}) and (\ref{cricurdisp0}) -- (\ref{cricurdisce}) 
yield the instability criterion of the two-stream instability
in the current sheet
\textcolor{black}{
\begin{eqnarray}
n_0 \delta^2 && <  \frac{8 m_{\rm e}}{\mu_0 e^2}
 \frac{1 + \xi/\gamma_0}{( \gamma_{\rm i} \gamma_{\rm e})^2}
\frac{1}{f^2}
+  \frac{4 m}{\mu_0 e^2} \sigma 
= 2.24 \times 10^{14} {\rm [m^{-1}]}
\left [
 \frac{1 + \xi/\gamma_0}{( \gamma_{\rm i} \gamma_{\rm e})^2} \frac{1}{f^2}
+  \frac{m}{2 m_{\rm e}} \sigma \right ] , 
\end{eqnarray}
where we define
$\displaystyle f \equiv \sqrt{\frac{\xi \Gamma_-}{\bar{M}_-/m_{\rm e}}} +
\sqrt{\frac{\Gamma_+ }{\bar{M}_+/m_{\rm e}}}$, 
$ \displaystyle \bar{M}_+ \equiv m_+ + \frac{\Gamma_+}{\Gamma_+ - 1} 
\frac{m}{2(1 +\xi / \gamma_0)} \sigma$, 
$ \displaystyle \bar{M}_-  \equiv m_- + \frac{\Gamma_-}{\Gamma_- - 1} 
\frac{\xi m}{2(1+ \xi / \gamma_0)} \sigma$.
}

\textcolor{black}{
In the case of the non-relativistic magnetic field limit, $\displaystyle 
\sigma \ll \frac{m_{\rm e}}{m_+} \ll 1$ and $\displaystyle \Gamma_- = \frac{5}{3}$, we have
\begin{equation}
n_0 \delta^2 < 1.34 \times 10^{14} \left ( 1 + \frac{1}{\xi} \right ) \; {\rm m^{-1}}.
\label{approsig1nonrela}
\end{equation}
In the case of a relativistic magnetic field, 
$\sigma \gg 1$, $\gamma_0 \gg 1$, and $\displaystyle \Gamma_\pm = \frac{4}{3}$,
we obtain
\begin{equation}
n_0 \delta^2 < 5.59  \times 10^{17}  \sigma \; {\rm m^{-1}}.
\label{approsig1ext}
\end{equation}
}
\textcolor{black}{Comparing Eqs. (\ref{approsig1nonrela}) and (\ref{approsig1ext}), 
we find that the critical value of $n_0 \delta^2$ in the case of the relativistically 
strong magnetic field 
is more than $1\,000$ times larger than
that in the case of a non-relativistic magnetic field.
This clearly demonstrates that special relativistic effects drastically enhance 
the two-stream instability.}

Using the parameters of the disk observed for Sgr A* by EHT \citep{eht22}:
$n_0 = 10^{12} \, \rm m^{-3}$, \textcolor{black}{$\xi =1$,
$\gamma_0 = \gamma_{\rm i} = \gamma_{\rm e} = 1$}, 
we obtain the stable condition to be $\delta > 10 \,$m 
\textcolor{black}{from Eq. (\ref{approsig1nonrela})}
and the threshold length of the instability to be
$l_{\rm th}^{\rm SgrA*} = 10 \,$m.
With respect to the AGN of M87 (M87*), using the parameters of EHT\cite{eht19} ---
$n_0 = 3 \times 10^{10} \, \rm m^{-3}$, $\xi =1$, and
$\gamma_0 = \gamma_{\rm i} = \gamma_{\rm e} = 1$ ---
the threshold length of the instability is given as $l_{\rm th}^{\rm M87*} = 100 \,$m
from Eq. (\ref{approsig1nonrela}).
Because the scales of the turbulence in the
accretion disks surrounding the black hole in Sgr A* and M87* are 
of the order of $\delta_{\rm SgrA*} = 10^{10} \,$ m 
and $\delta_{\rm M87*} =7 \times 10^{13} \,$ m, respectively, and 
the disks are free of the two-stream instability, these conditions appear to be reasonable:
$\delta_{\rm SgrA*} \gg l_{\rm th}^{\rm SgrA*}$ and
$\delta_{\rm M87*} \gg l_{\rm th}^{\rm M87*}$.
In the very low-density region, like the region at the footpoint of an AGN jet,
the two-stream instability is caused when the density becomes small enough, 
for example, if we consider the thickness of the current sheet to be
$\delta = 5.6 \times 10^5 \,$ m and $\sigma = 10$, 
the threshold density of the instability is given by $n_{\rm th}  = 10^4 \, \rm m^{-3}$ 
from Eq. (\ref{approsig1ext}).
When the density falls below this threshold, 
the two-stream instability occurs \textcolor{black}{in the current sheet with a large
current density between the strong antiparallel magnetic field regions} 
to disrupt the current sheet,
and magnetic field annihilation occurs.
\textcolor{black}{This magnetic field annihilation should trigger the
supply of plasma from outside of the region to the low-density region.}
This mechanism of plasma supply may work in the strong magnetic field region near the black hole
to form the relativistic jet in the AGN of M87.
On the other hand, no relativistic jet is observed in Sgr A*, while
a strong magnetic field region is expected to exist
near the black hole in the AGN of Sgr A*.\cite{eht22}
The difference between the AGNs of M87 and Sgr A* may
come from the direction of the strong magnetic field. That is, in the strong
magnetic field region of the AGN in Sgr A*, the magnetic field lines are aligned
in the same poloidal direction and no antiparallel magnetic field is formed.
In such a strong magnetic field region, a current sheet is not formed because of
the lack of an antiparallel magnetic field, and the two-stream instability is never triggered.
An accurate observation of the circular polarization of the radio wave
would be helpful to verify this model of the plasma supply mechanism
in the relativistic jet-forming regions of AGNs.

Due to the magnetorotational instability,
the magnetic field in an accretion disk is observed to be turbulent. 
A strong magnetic field around 
a spinning black hole is formed from the magnetic field in a disk.
It appears that an antiparallel magnetic field \textcolor{black}{would} arise
in an area with a strong magnetic field.
In the near future, numerical simulations beyond ideal GRMHD should be used to 
clarify the details of \textcolor{black}{the entire process of jet formation,
including the formation and annihilation of the antiparallel magnetic field}.
Such numerical simulations are also expected to reveal the difference between 
the magnetic field configurations in strong magnetic field regions with and without 
relativistic jets around
the black holes in M87* and Sgr A*.
\textcolor{black}{During the two-stream instability process in a realistic 
situation, the plasma would be non-uniform and magnetized. The dependence of the instability
on these additional factors is important and will be
clarified using the relativistic two-fluid equations including these factors in the near future.}

\begin{acknowledgments}
We are grateful to Mika Inda-Koide for her helpful comments on this paper.
\textcolor{black}{We also thank Seiji Ishiguro, Takayoshi Sano, and Hideo Sugama 
for their helpful discussions.}
\end{acknowledgments}

\appendix


\end{document}